\documentclass[10pt]{article}
\pdfoutput=1 

\usepackage{jcappub} 

\title{\boldmath\LARGE The fraction of muon tracks in cosmic neutrinos}

\author[a,b]{\large Francesco Vissani,}
\author[a,1]{\large Giulia Pagliaroli,\note{Corresponding author.}}
\author[c]{\large Francesco L. Villante}

\affiliation[a]{INFN, Laboratori Nazionali del Gran Sasso (LNGS), Assergi (AQ), Italy}
\affiliation[b]{Gran Sasso Science Institute (INFN), L'Aquila, Italy}
\affiliation[c]{Dipartimento di Scienze Fisiche e Chimiche, Universit\`a di L'Aquila, L'Aquila, Italy}

\emailAdd{francesco.vissani@lngs.infn.it}
\emailAdd{giulia.pagliaroli@lngs.infn.it}
\emailAdd{francesco.villante@lngs.infn.it}

\abstract{The study of the distintive signatures of the ultra high energy events recently seen by IceCube \cite{Aartsen:2013bka, talk,talk2,Koppes} can allow to single the neutrino origin out.  The detection of tau neutrinos would be a clear way to prove that they come from cosmic distances, but at the highest energies currently seen, about 1 PeV, an experimental characterization of tau events is difficult.
The study of the fraction of the muon tracks  seems more promising. 
In fact, for any initial composition, because of the occurrence of
flavor oscillations and despite their uncertainties, the fraction of
muon tracks in the cosmic neutrinos is smaller than the one of
atmospheric neutrinos, even hypothesizing an arbitrarily large
contribution from charmed mesons. A good understanding of the
detection efficiencies and the optimization of the analysis
  cuts, along with a reasonable increase in the statistics, should provide us with a significant test of the cosmic origin of these events.}

\begin{document}

\maketitle

\flushbottom

\section{Introduction}

The two events  announced by the IceCube collaboration
at Neutrino 2012 and recently 
described in  \cite{Aartsen:2013bka} are of enormous interest \cite{inte1,inte2,inte3,inte4,inte5,inte6,inte7,inte8,inte9,inte10,inte11} since
they could be the first indication of a high energy cosmic neutrino population.
It is unlikely that they are produced by atmospheric neutrinos
since the atmospheric neutrino flux is extremely low at PeV energy. 
Moreover, these events are detected as showers in the IceCube detector 
and thus they have been attributed to electron neutrino (or perhaps to neutral current) interactions \cite{Aartsen:2013bka};
by contrast, the  atmospheric neutrino flux is mostly composed by 
$\nu_\mu$ and $\overline{\nu}_\mu$. The absence of more events above PeV could suggest that these events come from a new 
population of neutrinos that is not power law distributed. Finally, it is exciting to note that a further set of 26 events 
has been preliminary announced \cite{talk}. These events 
could fill the gap at lower energies, till the region of the spectrum where 
ordinary atmospheric neutrinos dominate.
Note incidentally that a 2~$\sigma$ excess was already present in 
a muon sample $\sim 3$ times smaller than the present one, see Fig.4 of \cite{talk2}.
An interpretation of the IceCube observations in terms of  cosmic neutrinos is very attractive and the key 
 question becomes how to test this hypothesis. In this note, we argue that a viable possibility 
is to verify whether the fraction of muon neutrinos obeys the predictions of 3 flavor neutrino 
oscillations.\footnote{Three flavor neutrino oscillations have been widely discussed as a tool to measure the oscillation parameters, 
to learn on the original flavor composition of the neutrinos, to test the production mechanism or exotic scenarios, {\em etc.} \cite{b1,b2,b3,b4,b5,b6}; 
our goal here is to discuss the less ambitious but preliminary and urgent question,
concerning the origin of these events. The importance of the shower
events has been emphasized in \cite{b7}.}


\section{Flavor composition of  cosmic and atmospheric neutrinos}
For neutrinos travelling over cosmic distances, the oscillation probabilities are constant~\cite{gribov} and are given
 by $P_{\ell\ell'} = \sum_{\rm i=1,3} \, | U_{\ell \rm i}^2 |  \, | U_{\ell' \rm i}^2 |$ where $\ell,\ell'=e,\mu,\tau$ and $U$ 
is the neutrino mixing matrix.
By using the latest determinations of the oscillation parameters~\cite{lisinew}, 
updated after the measurement of $\theta_{13}$, we obtain that
the approximate numerical values of the probability matrix are:
\begin{equation}
P=
\left(
\begin{array}{ccc}
.548 & .244 & .208 \\
        & .404 & .352\\
        &         & .439
\end{array}
\right)
\label{P}
\end{equation}
Let us consider the fraction of neutrinos with given flavor, namely
\begin{equation}
\xi_\ell= \frac{\Phi_\ell}{\Phi_e+\Phi_\mu+\Phi_\tau}
\end{equation}
where $\Phi_\ell$ is the flux and $\ell=e,\mu$ or $\tau$.
The blend of neutrinos (or antineutrinos) at the source, that we generically indicate as $(\xi^0_e :\xi^0_{\mu} :\xi^0_{\tau})$,
is transformed in a mixture $(\xi_e : \xi_{\mu} : \xi_{\tau})$   of all neutrino types 
according to 
\begin{equation}
\xi_{\ell}  = \sum_{\ell'} P_{\ell\ell'} \, \xi^0_{\ell'}~.
\end{equation}
Let us assume, e.g. that neutrinos are produced by $\pi^+$ decay,
deriving from collisions of high energy protons with $\gamma$ rays in the vicinity of their source.
The original flavor composition is $(1:1:0)$  for neutrinos and $(0:1:0)$ for antineutrinos and, thus, we obtain 
$(\xi^0_e:\xi^0_{\mu}:\xi^0_{\tau})=(1/3 : 2/3 : 0)$ where we summed the neutrinos and antineutrinos and
 we considered that $\sum_{\ell} \xi_{\ell}= 1$.
After oscillations, this flavor composition becomes
\begin{equation}(\xi_e:\xi_{\mu}:\xi_{\tau}) =(0.35 : 0.35 : 0.30)\end{equation}
from which we see that a non negligible fraction of tau neutrinos is expected and
the muon neutrino fraction $\xi_\mu$ is reduced to values that are lower than 0.5.

\begin{figure}[t]
\par
\begin{center}
\includegraphics[width=.32\textwidth,angle=0]{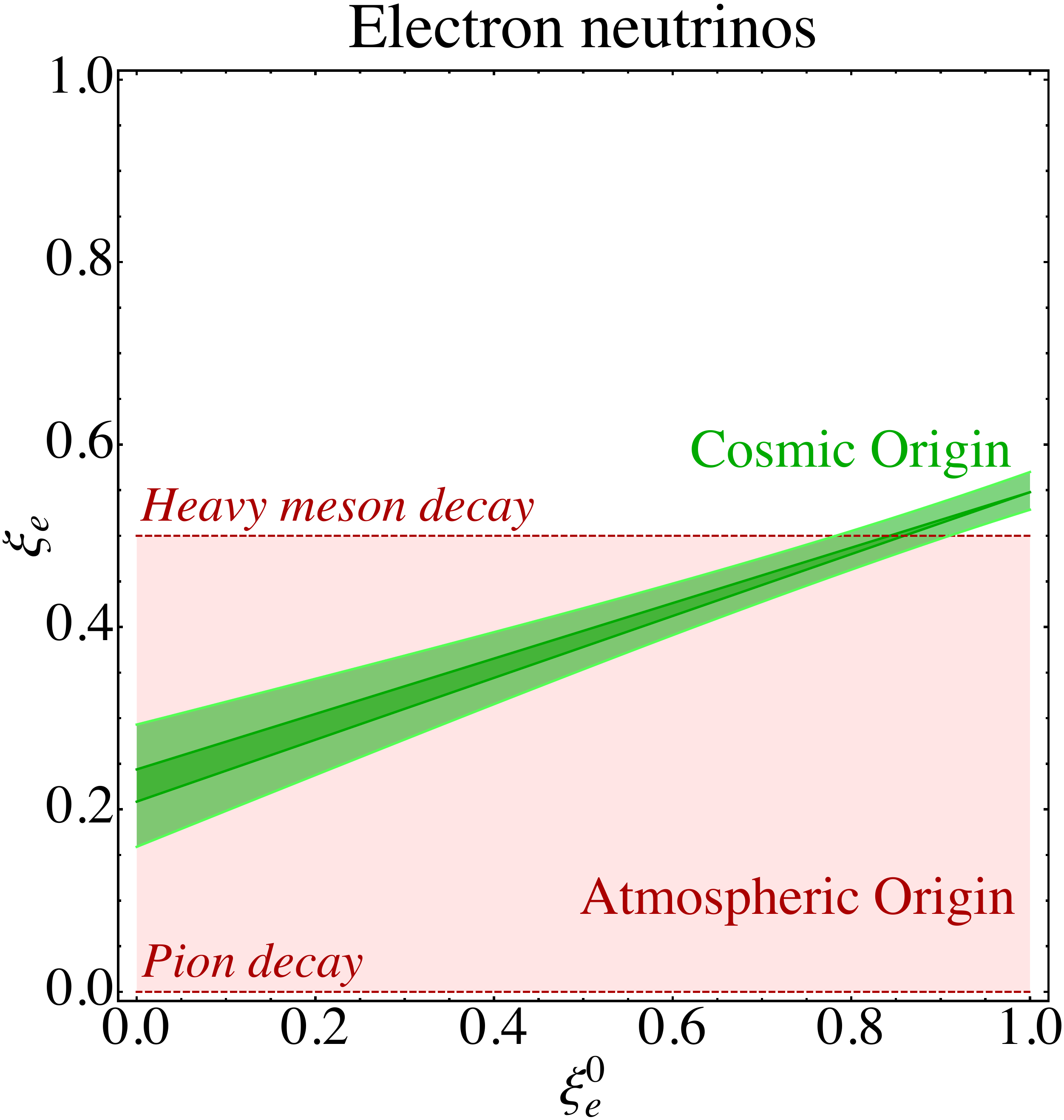}
\includegraphics[width=.32\textwidth,angle=0]{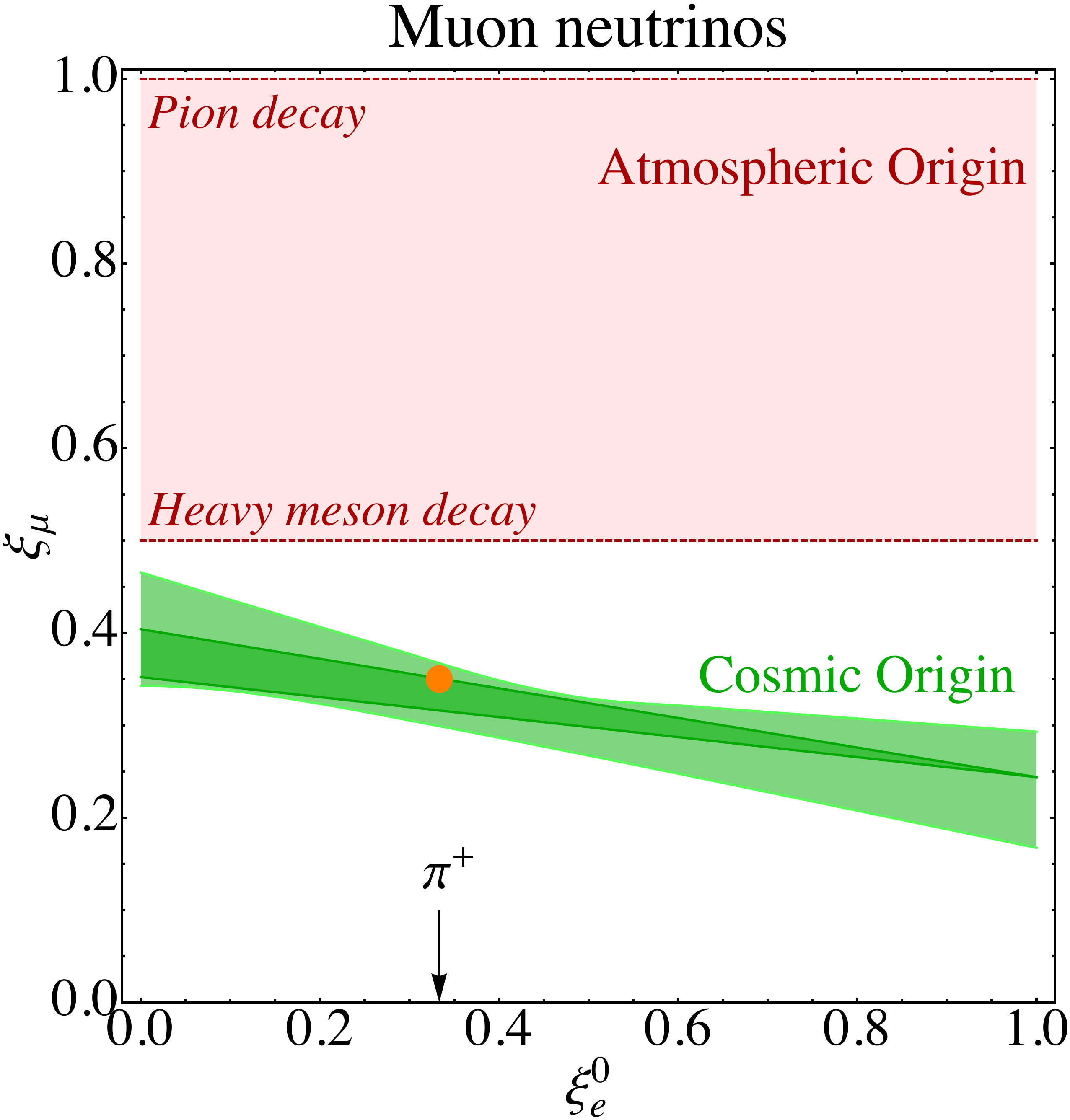}
\includegraphics[width=.32\textwidth,angle=0]{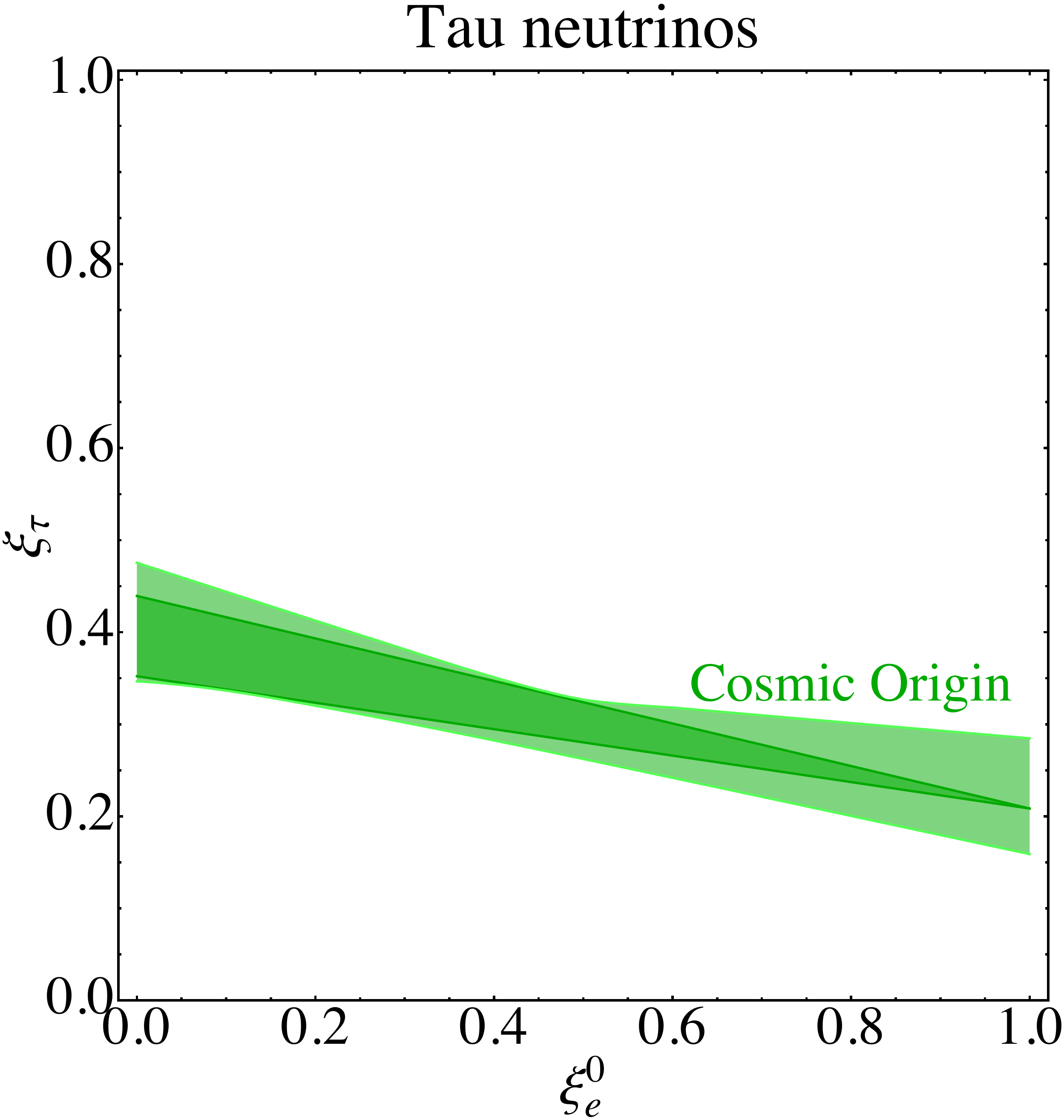}
\end{center}
\par
\vspace{-5mm} \caption{\em\protect\small The flavor composition of
  cosmic and atmospheric neutrinos. The dark green regions show the
  spread due to different initial flavor composition of the
  cosmic component. The light green bands represent the additional spread
  due to the uncertainty on the oscillation parameters.}
\label{fig1}
\end{figure}

We will show that the above conclusions are generic since they are essentially 
independent on the original composition of the neutrino flux.
In fact, assuming the oscillation probabilities reported in Eq.~(\ref{P}), 
the arrival neutrino flavor fractions are given by:
\begin{eqnarray}
\nonumber 
\xi_e &=& 0.244 + 0.304\, \xi^0_e - 0.035\, \xi^0_\tau \\
\nonumber
\xi_\mu &=& 0.404 - 0.160\, \xi^0_e - 0.052\, \xi^0_\tau   \\
\xi_\tau &=& 0.352 - 0.144\, \xi^0_e + 0.087\, \xi^0_\tau
\label{ratios}
\end{eqnarray}
where we considered that, whichever is the neutrino production mechanism,  $\xi^0_{\mu} = 1 -\xi^0_{e}-\xi^0_{\tau}$.
The dark green areas in Fig.~\ref{fig1} show the flavor ratios predicted by Eqs.~(\ref{ratios}) 
for an arbitrary value of the initial electron neutrino fraction $\xi^0_e$ when 
$\xi^0_{\tau}$ is varied in the physical range $0\le\xi^0_{\tau} \le 1-\xi_e^0$.
In general terms, we have that, for any fixed $\xi_e^0$,  
the fraction $\xi_\ell$ of cosmic neutrinos with flavor $\ell=e,\mu,\tau$ 
that reach us is comprised between:
\begin{equation}
P_{\mu \ell}+ (P_{e\ell}- P_{\mu \ell})\, \xi_e^0
\; \;  \; \mbox{   and   }  \; \;  \; 
P_{\tau \ell}+ (P_{e\ell}- P_{\tau \ell})\, \xi_e^0
\label{limits}
\end{equation}
where the first limit corresponds to the assumption
 $\xi_{\tau}^0 = 0$ while the second
is obtained by assuming $\xi_{\mu}^0=0$ (i.e., $\xi_{\tau}^0 = 1 -\xi_{e}^0$);
this range is narrow 
since $P_{\mu \ell}\sim P_{\tau \ell}$ as it is seen
from Eq.~(\ref{P}). 
Also the dependence of $\xi_\mu$ and 
$\xi_\tau$ on $\xi_e^0$ is not very strong, 
due to partial cancellations between $P_{\mu\ell}$, $P_{\tau \ell}$ and 
$P_{e \ell}$. This holds true even considering  the cases when the 
initial fraction of electrons $\xi_e^0$ and of tau $\xi_\tau^0$ are 
energy dependent.

\begin{figure}[t]
\par
\begin{center}
\includegraphics[width=.6\textwidth,angle=0]{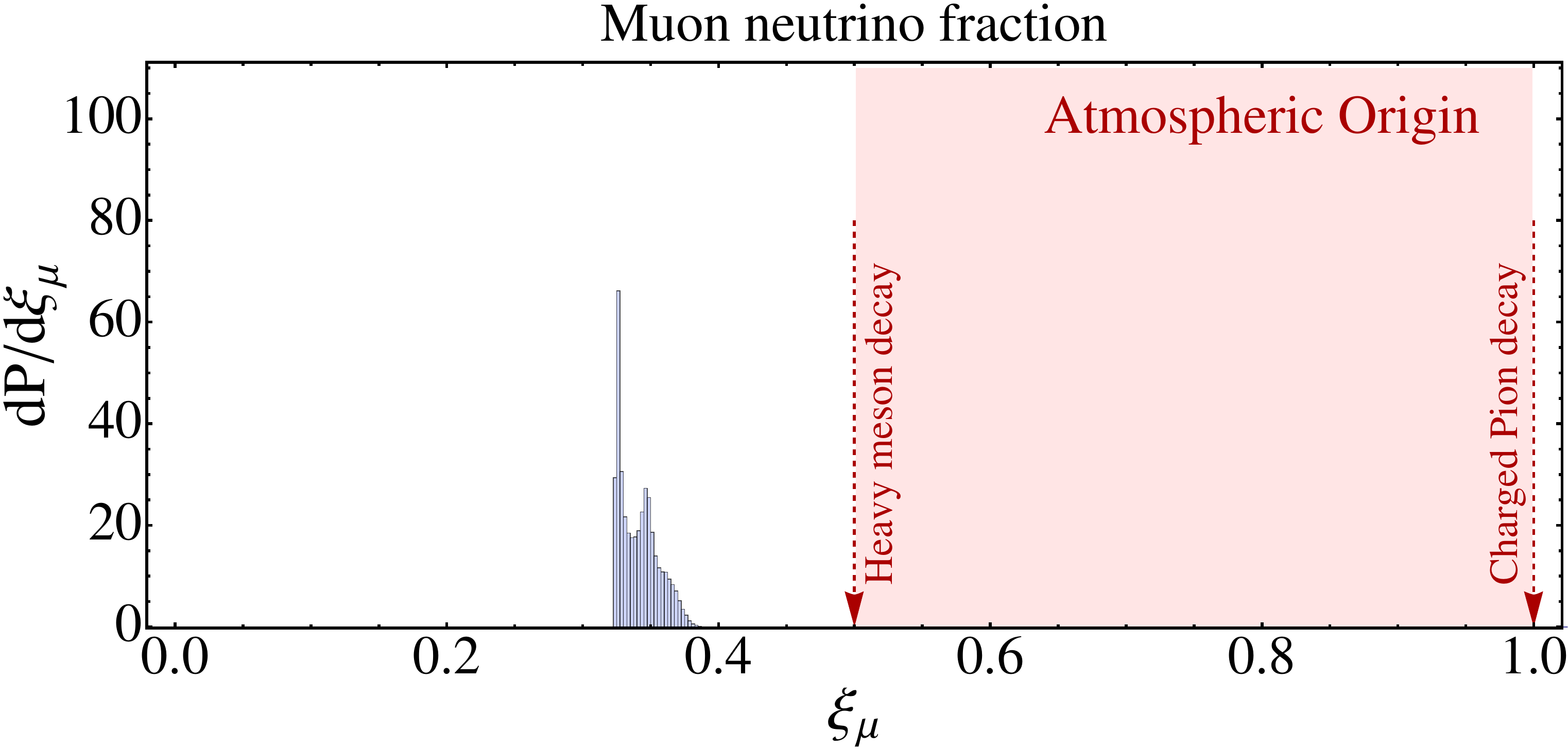}
\end{center}
\par
\vspace{-5mm} \caption{\em\protect\small The probability distribution of $\xi_\mu$ for cosmic neutrinos 
produced by charged pion decay obtained by propagating the uncertainties 
in neutrino oscillation parameters.}
\label{fig2}
\end{figure}

Our considerations are not affected by the uncertainties in neutrino oscillation parameters.
To quantify their relevance, we constructed the likelihood distributions of
$\sin^2 \theta_{12} $, $\sin^2 \theta_{13} $ and $\sin^2 \theta_{23} $
from the $\chi^2$ profiles given by \cite{lisinew} (assuming negligible correlations).
For the CP-violating phase $\delta$, in order to be extremely 
conservative, we took a flat distribution between $\left[0,2\pi\right]$. 
We then determined the probability distributions of $\xi_e$, $\xi_\mu$ and $\xi_\tau$ 
by MonteCarlo extraction of the oscillation parameters.
In Fig.~\ref{fig2} we show the result obtained for the specific case of cosmic neutrinos
produced by charged pions. The muon neutrino fraction $\xi_\mu$ derived assuming
the best-fit values for oscillations parameters in \cite{lisinew} is $\xi_\mu =0.35$. By propagating the
uncertainties, we obtain a $\xi_\mu$ probability distribution which has a finite width and
a non trivial structure resulting from the non linear dependence of the $P_{\ell\ell'}$ 
from the oscillation parameters (see \cite{fvx,fx} and \cite{next} for a discussion
of this aspect). However, the conclusion that  the expected muonic neutrino fraction at Earth is reduced to 
values that are lower than 0.5 is not spoiled. In fact, the $90\%$ C.L. allowed range is 
\begin{equation}
\xi_\mu =0.325-0.37,
\end{equation}
as it is obtained by integrating out symmetrically 
the $5\%$ on both sides of the distribution.
A general appraisal of the relevance of oscillation parameter uncertainties is finally
obtained from the light green areas in Fig.~\ref{fig1} that show the spread of the limits in Eq.~(\ref{limits}) 
when 90$\%$ C.L. allowed ranges are considered.

The green regions in Fig.~\ref{fig1}  should be compared with expectations for the atmospheric neutrino component.
At high energies, atmospheric electron neutrinos are rare  
since the muons produced in the atmosphere have no 
time to decay (see e.g., \cite{ginz,gaisser})
and oscillations are not effective.\footnote{Let us consider neutrinos with energies between 10 TeV and 10 PeV. The phases of 3 flavor 
oscillations, $\Delta m^2  L c^3/ (4 E \hbar)$, are smaller than 1/300 
for atmospheric neutrinos ($L<2 R_\oplus$) and larger than 400 already 
for neutrinos coming from Proxima Centauri ($L=4.2$ ly): therefore, for the energies of interest, vacuum 
oscillations are fully developed for cosmic neutrinos and absent for atmospheric neutrinos.} 
As a consequence, the conventional production mechanism through charged pions gives a flavor composition $\sim (0:1:0)$. 
The production of heavy (charmed) mesons introduces an additional component, the so-called `prompt' atmospheric neutrinos, 
with a flavor composition $(1/2 : 1/2 : 0)$: see e.g., \cite{ml}. The
prompt neutrino component contains electron neutrinos, 
thus it can account for the observed shower events, 
and the electron-to-muon neutrino ratio is close to the one expected from cosmic sources. 
The number of prompt neutrino events is estimated to be small 
and marginally observable with the present statistics;
however,  in view of the large theoretical uncertainty, it is worthwhile to consider a 
conservative attitude, and to assume that the atmospheric flavor
composition is between the two extreme situations represented by
charged pions and charmed mesons decay. 
Such a flavor composition of the atmospheric neutrino events is
represented in Fig.\ref{fig1} by the red horizontal bands and in
Fig.\ref{fig2} by the red vertical band.

From this conservative standpoint, the observation of shower events of ultra-high energies is not sufficient to discriminate 
between cosmic and atmospheric origin of the events--as it is seen in the left panel of Fig.\ref{fig1}--even if the mere existence of showers at PeV
shows that there is something more than  the conventional atmospheric neutrinos \cite{b7}. 
The crucial information is instead 
provided by the observations of tau neutrinos of ultra-high energies 
(absent in the conventional atmospheric neutrinos) and/or by the 
determination of the muon neutrino fraction $\xi_\mu$ around PeV,
as we discuss in the following.


\section{Testing the cosmic origin}
We discuss a couple of tests to address the cosmic origin of high energy events observed in neutrino telescopes.
The most direct evidence could be provided from the observation of tau neutrinos, by seeing some manifestation of the two vertices
 where tau is produced and where it decays. 
However, the efficiencies for reliable identification of taus are not very large, especially for energies of 1 PeV or lower. 
In fact, the separation among the strings is about 120 m. If we ask that the two vertices are separated by this distance, and equate it to the theoretical expression  
$E/(m c^2)\times (c \, t_{1/2})$ (where $E$, $m$ and $t_{1/2}$ are the energy, mass and half-life of the tau), we get that the energy should be $E=3.5$ PeV, see e.g. \cite{b1}. 
This is a region of energies where no neutrino has been seen yet. If
this were due to a cut in the spectrum then the search for tau events
would be challenging. In this case, we should rely on fluctuations, very refined analyses and/or new setup, 
before we can firmly claim the presence of a tau event.
(For more discussion, see the paper of IceCube concerning the search of tau neutrino, in the range of energies between 0.3 and 200 PeV \cite{tau}.)

An alternative test, that should be possible below PeV
(allowing us to increase the statistics)
is to show that the muon neutrino fraction $\xi_{\mu}$ of the observed population 
does not fall in the range $(0.5,1)$ expected for atmospheric neutrinos.
This relies on the possibility to identify $\nu_\mu$ charged current interactions  in the detector 
by looking at the presence of muon tracks.
The rate  $R_{\rm T}$ of track events produced by a flux $\Phi_\mu$ of
muon neutrinos is estimated as: 
\begin{equation}
R_{\rm T} = N_{\mbox{\tiny tar}} \varepsilon_{\rm T} \int_{\overline{E}} dE_{\nu} \, \sigma_{\rm CC}(E_{\nu}) \, \Phi_\mu(E_{\nu})
\label{tracks}
\end{equation}
where $N_{\mbox{\tiny tar}}$ is the number of target nuclei, $\sigma_{\rm CC}$ is the charged current cross 
section and $\varepsilon_{\rm T}$ is the efficiency of $\nu_\mu$ events identification (which includes the detector volume 
reduction implied by analysis cuts). 
The rate $R_{\rm S,CC}$ of shower events produced by charged current interactions of
electron and tau neutrinos is   
\begin{equation}
R_{\rm S, CC} = N_{\mbox{\tiny tar}}  \varepsilon_{\rm S} \int_{\overline{E}} dE_{\nu} \, \sigma_{\rm CC}(E_{\nu}) \, \left[ \Phi_e (E_{\nu})+\Phi_{\tau}(E_{\nu})\right]
\label{showers}
\end{equation}
where $\varepsilon_{\rm S}$ is the shower detection efficiency.
The fraction of track events is, thus, equal to
\begin{equation}
f\equiv\frac{R_{\rm T}}{R_{\rm T}+R_{\rm S, CC}} = \frac{\eta \, \xi_{\mu} }{ 1 -\xi_{\mu} \left( 1- \eta \right)}
\label{trackratio}
\end{equation}
where $\eta = \varepsilon_{\rm T}/\varepsilon_{\rm S}$ is the ratio of efficiencies and we considered that charged current
cross sections of different neutrino flavors are approximately equal.
In the above relation, we implicitily assumed that showers and tracks probe the
same portion of the neutrino spectrum. In other words the rates (\ref{tracks}) and (\ref{showers}) are 
obtained by integrating the fluxes of the different neutrino
flavors above the same energy threshold ${\overline E}$.\footnote{In principle, this requirement is not strictly necessary but
it permits to obtain predictions which are essentially independent on the theoretical assumptions 
about the neutrino spectrum. The analysis thresholds, which can be eventually different for tracks 
and showers, have to be optimized with the detailed knowledge of the
experimental apparatus.}
This implies that the reconstruction of the initial state neutrino energy is good enough. 
For shower events, the energy reconstruction is accurate at the $\sim 15\%$ level \cite{talk}. 
The accuracy is much worse for a generic track event 
which may be produced by a $\nu_\mu$ interacting outside the detector. 
For this reason, it is better to consider only contained vertex track events in which the interaction point 
of the $\nu_\mu$ is visible. In this case, in fact, the $\nu_\mu$ interaction with the nucleon 
produces an hadronic cascade that deposits its energy inside the 
sensitive volume and a muon whose energy can be reconstructed 
from the rate of catastrophic energy losses along the track produced in the detector
\cite{icrc2011}. 
We note that the recent IceCube analysis \cite{Koppes} 
includes only contained vertex events of highest energy,
with the goal of reducing the background due to atmospheric muons and neutrinos.

Few sub-dominant effects can modify the simple estimate given above.
First, we neglected neutral current contribution to shower event rate. 
This is expected to be small because the energy deposited in the 
detector, in the form of an hadronic shower, is a limited fraction of the initial neutrino energy. 
Most of the energy is, in fact,  carried away by the final state neutrino.
When $E_{\nu}=1-10$ PeV, we have that $\langle E_{\mbox{\tiny hadr}}\rangle\approx (1/4) \, E_\nu$, 
indicating that, an event with 1 PeV hadronic energy is produced
by neutrinos with average energy of about 4 PeV.
By taking this into account, the rate $R_{\rm S, NC}$ of neutral current events is estimated as:
\begin{equation}
R_{\rm S, NC}  = N_{\mbox{\tiny tar}} \varepsilon_{\rm S} \int_{4\overline{E}} dE_{\nu} \, \sigma_{\rm NC}(E_{\nu}) \, \left[ \Phi_{e}(E_{\nu}) +\Phi_\mu(E_{\nu}) +\Phi_\tau(E_{\nu})\right] 
\end{equation}
If we postulate conservatively that the flux diminishes as $\Phi\propto E_\nu^{-2}$
we obtain:
\begin{equation}
\frac{R_{\rm S, NC} }{R_{\rm S, CC}}\approx A_{\rm NC} \, \frac{1}{1-\xi_\mu}
\end{equation}
with $A_{\rm NC}=0.17$.\footnote{
The coefficient $A_{\rm NC}$ is reduced to $0.13$ if the slope of
the spectrum, $\Phi\propto E^{-\alpha}$, is equal to $\alpha=2.2$,
whereas $A_{\rm NC}=0.11$ if there is an
exponential cut at energies 20 times more than the threshold (e.g.,
$\bar{E}=100$ TeV and $E_{\mbox{\tiny cut}}=2$ PeV).} In the above relation, we have taken into
account that $\sigma_{\rm NC}/\sigma_{\rm CC}\sim 0.38$ almost independently on energy, and that the cross sections $\sigma_{\rm NC}$ amd $\sigma_{\rm CC}$ scale approximately as $E_{\nu}^{0.44}$ \cite{Gandhi}.
As an additional remark,  we note that in the case of $\nu_\tau$ CC-interactions, a fraction $\sim 20\%$ 
of the initial neutrino energy is not detectable since it is lost in neutrinos produced by tau decay 
\cite{EnergyLost}. 
By using the arguments above, it can be calculated 
that this corresponds to reducing the $\nu_\tau$ signal by a coefficient $A_{\tau} \approx 12\%$.
If we take the above effects into account, Eq. (\ref{trackratio}) is slightly modified. We obtain: 
\begin{equation}
f\equiv\frac{R_{\rm T}}{R_{\rm T}+R_{\rm S, CC}+R_{\rm S, NC}} = 
\frac{\eta \, \xi_{\mu} }{ (1+A_{\rm NC})  - A_{\tau} \,\xi_{\tau} - \xi_{\mu} \left( 1- \eta \right)}
\label{trackratio2}
\end{equation}
that will be used in the next section to evaluate the perspectives of the proposed measure.


\section{Perspectives}
In conclusion, a possible test of the cosmic origin of the events
observed by IceCube is simply to measure the fraction of muon tracks $f$ precisely enough.
In fact, considering for simplicity the case of cosmic neutrinos
with flavor composition $(\xi_e,\xi_\mu,\xi_\tau) = (1/3,\,1/3,\,1/3)$
and assuming $\eta=1$, we obtain from Eq.(\ref{trackratio2}):
\begin{equation}
f= \left\{
\begin{array}{cc}
 0.29 & \mbox{ for cosmic neutrinos}\\[2ex]
{}> 0.43 & \mbox{ for atmospheric neutrinos}
\end{array}
\right.
\end{equation} 
Note that the case $f=0.43$ corresponds to assuming that the
atmospheric neutrino flux 
is solely due to charm mesons decays (prompt contribution).

In order to illustrate the potential of this measure, 
simple statistical considerations are sufficient. 
Let us consider the normalized binomial distribution $B(n|f,N)=
{\small (\!\!\begin{array}{c}N\\[-1ex] n\end{array}\!\!)} 
\, f^n\, (1-f)^{N-n}$ where the 
true frequency is $f$ and $N$ is the total number of events.
The possibility that the frequency is $f'>f$ can be excluded at a certain confidence level CL 
requires that $N$ satisfies the condition 
\begin{equation}
1-\mbox{CL}>\sum_{n>f' N}^N B(n|f,N)
\end{equation}
We assume the cosmic origin of the events ($f=0.29$) and, by using the above expression, we conclude that 
the charm hypothesis ($f'=0.43$)  can be excluded as an alternative explanation when the total number 
of events is larger than:
\begin{equation}
N=73,\ 125 \mbox{ at a CL=} 99,\ 99.9\%, \mbox{ respectively}.
\end{equation}

The assumption $\eta=1$ corresponds to the optimal situation in which the shower and track detection 
efficiencies are equal. If contained vertex events are only considered,  this assumption is, in principle, adequate
since different neutrino flavors have the same interaction
volume. However, the analisys cuts needed to reduce the background
unavoidably modify the value of $\eta$. 
As an example, by imposing a fixed threshold for the energy released in the detector,
a stronger selection of track events is performed. The muon range is, in fact,  larger than 
the size of the apparatus, at $\sim 1~{\rm PeV}$ is $\sim 15~{\rm km}$
and, thus, muons deposit only a fraction of their energy in the detector.  
The value of $\eta$ is consequently reduced as it is also seen by
comparing the effective areas for $\nu_e$, $\nu_\mu$ and $\nu_\tau$ detection
in the recent IceCube analysis reported by \cite{Koppes}.  
To evaluate the importance of detection efficiencies, we repeat our analysis for  $\eta =0.5$. In this assumption,
Eq.~(\ref{trackratio2}) gives $f=0.17$ for the cosmic origin and $f' = 0.27$ for the charm hypothesis. To discriminate 
between the two options, the total number of events needs to be larger than:
\begin{equation}
N=96,\ 164 \mbox{ at a CL=} 99,\ 99.9\%, \mbox{ respectively}.
\end{equation}

In the above simplified considerations, we neglected 
the contamination of atmospheric neutrinos and 
few additional effects,  such as e.g.  the $\nu_\tau$ regeneration in the earth 
interior \cite{Montaruli}, etc.,  that  could modify the expected
value of $f$ and, thus, affect the estimate of the required statistics.
However, the general picture seems promising. In summary, the existing statistics has to be increased 
by a small multiplicative factor ($\sim {\rm few}$) whose magnitude depends on the specific cuts adopted in 
the experimental data selection procedure.
The IceCube collaboration has the data and the tools to optimize the cuts and
perform a complete analysis: we hope that they will shed light on the origin of these exciting events.

\end{document}